\documentclass[11pt,aps,nofootinbib,prd,aps,epsf,floats,axodraw,
               amsmath,amssymb,amsfonts]{revtex4}
\usepackage{amsmath, amssymb}
\newcommand{\mathsym}[1]{{}}

\usepackage{graphicx}
\usepackage{amsmath}
\usepackage{amssymb}
\usepackage{bm}
\newcommand{\be}{\begin{equation}}
\newcommand{\ee}{\end{equation}}
\newcommand{\bea}{\begin{eqnarray}}
\newcommand{\eea}{\end{eqnarray}}

\newcommand{\rem}[1]{}
\newsavebox{\PSLASH}
 \sbox{\PSLASH}{$p$\hspace{-1.8mm}/}
 
\renewcommand{\theequation}{\thesection.\arabic{equation}}
\newcounter{saveeqn}
\newcommand{\add}{\addtocounter{equation}{1}}
\newcommand{\alpheqn}{\setcounter{saveeqn}{\value{equation}}%
\setcounter{equation}{0}%
\renewcommand{\theequation}{\mbox{\thesection.\arabic{saveeqn}{\alph{equation}}}}}
\newcommand{\reseteqn}{\setcounter{equation}{\value{saveeqn}}%
\renewcommand{\theequation}{\thesection.\arabic{equation}}}

 \newsavebox{\notrightarrow}
 \sbox{\notrightarrow}{$\to$\hspace{-4mm}/}
 
 \newsavebox{\PARTIALSLASH}
 \sbox{\PARTIALSLASH}{$\partial$\hspace{-1.6mm}/}
 
 \newsavebox{\ASLASH}
 \sbox{\ASLASH}{$A$\hspace{-2.1mm}/}
 
 \newsavebox{\KSLASH}
 \sbox{\KSLASH}{$k$\hspace{-1.8mm}/}
 
 \newsavebox{\LSLASH}
 \sbox{\LSLASH}{$\ell$\hspace{-1.8mm}/}
 
 \newsavebox{\QSLASH}
 \sbox{\QSLASH}{$q$\hspace{-1.8mm}/}
 
 \newsavebox{\DSLASH}
 \sbox{\DSLASH}{$D$\hspace{-2.2mm}/}
 
 \newsavebox{\DbfSLASH}
 \sbox{\DbfSLASH}{${\mathbf D}$\hspace{-2.8mm}/}
 
 \newsavebox{\DELVECRIGHT}
 \sbox{\DELVECRIGHT}{$\stackrel{\rightarrow}{\partial}$}
 
 \newcommand{\blue}{\IfColor{\textCadetBlue}{}}
\newcommand{\black}{\IfColor{\textBlack}{}}
\newcommand{\red}{\IfColor{\textRed}{}}
\newcommand{\green}{\IfColor{\textOliveGreen}{}}
\newcommand{\lila}{\IfColor{\textRedViolet}{}}

\begin{document}
\begin{flushright}
 [math-ph]
\end{flushright}
\title{Groenewold-Moyal Product, $\alpha^*$-Cohomology,
and
Classification of
Translation-Invariant Non-Commutative Structures
}

\author{Amir Abbass Varshovi}\email{amirabbassv@ipm.ir}

\affiliation{
   School of Mathematics, Institute for Research in Fundamental Sciences (IPM).\\
   School of Physics, Institute for Research in Fundamental Sciences (IPM).\\
                                     Tehran-IRAN}
\begin{abstract}
       \textbf{Abstract\textbf{:}} The theory of $\alpha^{*}$-cohomology is studied thoroughly and it is shown that in each cohomology class there exists a unique 2-cocycle, the harmonic form, which generates a particular Groenewold-Moyal star product. This leads to an algebraic classification of translation-invariant non-commutative structures and shows that any general translation-invariant non-commutative quantum field theory is physically equivalent to a Groenewold-Moyal non-commutative quantum field theory.
\end{abstract}
 \maketitle
\section{Introduction}\label{introduction}
\indent  Translation-invariant star products as the most natural generalization of non-commutative Groenewold-Moyal $\star$ product have been introduced \cite{Lizzi,Galluccio} and discussed partially \cite{Vitale, Ardalan, Varshovilong, Varshovijmp, Varshovinew} in the framework of non-commutative quantum field theories by generally considering the pathological behavior of UV/IR mixing through the renormalization programs. In fact the most prominent motivation of translation-invariant products was essentially inspired by considering the Wick-Voros formalism \cite{Voros} in deformation quantization approach \cite{Lizzi, Galluccio, Vitale, Ardalan, Lizzi2, Saha, Basu} as an alternative technique of quantization for Groenewold-Moyal star product. But on the other hand, it was then shown \cite{Lizzi, Galluccio, Vitale, Lizzi2, Saha, Basu} that Wick-Voros non-commutative field theories are physically exactly equivalent to Groenewold-Moyal ones at all quantum levels. Particularly Wick-Voros and Groenewold-Moyal star products lead precisely to the same Green functions and consequently by LSZ theorem result exactly in the same scattering matrix for respectively the Wick-Voros and the Groenewold-Moyal non-commutative versions of any given renormalizable quantum field theory \cite{Lizzi, Galluccio, Varshovilong, Varshovinew}.\\
 \par The equivalence of non-commutative Wick-Voros and Groenewold-Moyal quantum field theories was then understood to be intimately correlated to a cohomology theory, so called $\alpha^*$-cohomology \cite{Lizzi, Galluccio, Vitale, Varshovilong, Varshovinew}, as an algebraic theory for classifying translation-invariant star products thoroughly inspired by the Hochschild theory of cohomology. In fact, by definition two translation-invariant complex \footnote {The star product $\star$ is said to be complex if; $(f \star g)^*=g^* \star f^*$ for any $f,g\in C^\infty (\mathbb{R}^m )$.} star products† $\star_1$ and $\star_2$ on $C^\infty (\mathbb{R}^m )$ are $\alpha^*$-cohomologous if and only if there exists a fixed smooth function over $\mathbb{R}^m$, say $\beta$, such that for any $n\geq1$, and any set of $\{f_i\}^n_1\subset C^\infty (\mathbb{R}^m )$ one finds that;
\begin{eqnarray}\label {1}
\int_{\mathbb{R}^m} f'_1\star_1 ... \star_1 f'_n =\int_{\mathbb{R}^m} f_1\star_2 ... \star_2 f_n~,
\end{eqnarray}
\noindent for:
\begin{eqnarray} \label {2}
f'(x)=\int \frac{\emph{\emph{d}}^m p}{(2\pi)^m} e^{ip.x} \tilde{f}(p) e^{\beta(p)}~,
\end{eqnarray}
\noindent $f\in C^\infty (\mathbb{R}^m )$, and for $\tilde{f}$ the Fourier transform of $f$ \cite{Varshovinew}.\\
\par It is seen that (\ref {1}) leads to the most general classification of $\star $ products from the viewpoints of quantum physics, so called the quantum equivalence [7]. By definition, two star products $\star_1$ and $\star_2$ are quantum equivalent if and only if there exists a fixed $\beta \in C^\infty (\mathbb{R}^m )$, with $\beta(0)=0$, such that for any $n \geq 1$, the equality
\begin{eqnarray} \label {3}
\tilde{G}_{\star_1~\emph{\emph{conn.}}}(p_1,...,p_n)=e^{\sum_{i=1}^n \beta(p_i)}~\tilde{G}_{\star_2~\emph{\emph{conn.}}}(p_1,...,p_n)
\end{eqnarray}
\noindent holds for any given renormalizable quantum field theory, where $G_\emph{\emph{conn.}}$ is any connected $n$-point function, $G_{\star~\emph{\emph{conn.}}}$ is its non-commutative version for the star product $\star$ and $\tilde{G}_{\star~\emph{\emph{conn.}}}(p_1,...,p_n)$ is its Fourier transform for the modes $\{p_i\}_{i=1}^n$.\\


 \par Therefore it is seen [7] that the star products $\star_1$ and $\star_2$ are $\alpha^*$-cohomologous if and only if they are quantum equivalent. Consequently, all the quantum behaviors of two $\alpha^*$-cohomologous non-commutative translation-invariant versions of a fixed given quantum field theory are exactly the same. This result can be considered as an algebraic proof for physical equivalence of Wick-Voros and Groenewold-Moyal non-commutative field theories provided these star products are $\alpha^*$-cohomologous.\\


 \par In this article translation-invariant products are studied in the framework of $\alpha^*$-cohomology theory. It is precisely shown that for any complex translation-invariant product $\star$ there exists a particular Groenewold-Moyal star product that is $\alpha^*$-cohomologous to $\star$. This eventually can be considered as a general version of the theorem which equalizes the Wick-Voros and the Groenewold-Moyal star products from the viewpoints of quantum physics. It is then strictly concluded that the non-commutative structure of space-time given by the commutation relation of coordinate functions via the star product, entirely characterizes the structure of abnormal quantum behaviors of non-commutative quantum field theories such as the structure of UV/IR mixings. While this correlation of the non-commutative structures of space-time and abnormal quantum behaviors was intuitively conjectured [12] and partially proved \cite{Lizzi, Galluccio, Vitale, Lizzi2} only for Wick-Voros and Groenewol-Moyal formalisms, but here it is accurately provided a strict algebraic proof for all general cases.\\


 \par In section II, $\alpha^*$-cohomology theory and its Hodge theorem are introduced and discussed referring to [7]. In section III it is shown that the harmonic forms due to the Hodge theorem, lead precisely to Groenewold-Moyal star products. Some algebraic theorems are also worked out in the following.\\


\par
\section{$\alpha^*$-Cohomology and the Hodge Theorem}
\setcounter{equation}{0}
\par As a definition a translation-invariant star product on $\mathbb{R}^m$ with respect to coordinate system $\{x^i \}_{i=1}^m$ is an associative multiplication over $C^\infty (\mathbb{R}^m )$ which manifestly doesn't depend on the coordinate functions $x^i$, $i=1,...,m$. Hence, more precisely, from the physical viewpoints a translation-invariant star product preserves the behavior of any Lagrangian density under translation when it is used instead of the ordinary product. Consequently translation-invariant star products lead to the energy-momentum conservation law for any relativistic quantum filed theory. Strictly speaking star product $\star$ on $C^\infty (\mathbb{R}^m )$ is translation-invariant if;
\begin{eqnarray} \label {4}
\mathfrak{T}_a(f) \star \mathfrak{T}_a(g)=\mathfrak{T}_a(f \star g)~,
\end{eqnarray}
\noindent for any vector $a\in{\mathbb{R}^m}$ and for any $f,g\in C^\infty (\mathbb{R}^m )$, where $\mathfrak{T}_a$, is the translating operator; $\mathfrak{T}_a (f)(x)=f(x+a)$, $f\in C^\infty (\mathbb{R}^m )$. Replacing $a$ with $ta$, $t\in R$, in (\ref {4}) and differentiating with respect to $t$ at $t=0$, the most important property of translation-invariant products, the exactness, is inferred;
\begin{eqnarray} \label {5}
\partial_\mu (f \star g)=\partial_\mu (f) \star g + f \star \partial_\mu (g)~,
\end{eqnarray}
\noindent $\mu=1,...,m$. The exactness property shows that the star product $\star$ as a function is not given in terms of the coordinate functions of which are intrinsically encoded in the translation operator $\mathfrak{T}$.\\


 \par An equivalent simple definition of translation-invariant products over the Cartesian space $\mathbb{R}^m$, is given by [1];
\begin{eqnarray} \label {6}
(f \star g)(x):= \int \frac{\emph{\emph{d}}^m p}{(2\pi)^m} \frac{\emph{\emph{d}}^m q}{(2\pi)^m} \tilde{f}(q) \tilde{g}(p) e^{\alpha (p+q,q)} e^{i(p+q).x}~,
\end{eqnarray}
\noindent for $f,g \in C^\infty (\mathbb{R}^m)$, their Fourier transformations $\tilde{f},\tilde{g} \in C^\infty (\mathbb{R}^m)$, and finally for a 2-cocycle $\alpha\in C^\infty (\mathbb{R}^m \times \mathbb{R}^m)$ which obeys the following cyclic property;
\begin{eqnarray} \label {7}
\alpha(p,q)+\alpha(q,r)=\alpha(p,r)+\alpha(p-r,q-r)~,
\end{eqnarray}
\noindent for any $p,q,r\in \mathbb{R}^m$. It is seen that (\ref {7}) holds if and only if $\star$ is associative, i.e.; $(f \star g)\star h=f\star (g \star h)$ for $f,g,h\in C^\infty (\mathbb{R}^m)$.\\


 \par To have a well-defined definition for translation-invariant products, $C^\infty (\mathbb{R}^m)$ is conventionally replaced by $\mathcal{S}_c (\mathbb{R}^m)$, the Schwartz class functions with compactly supported Fourier transforms [7]. On the other hand $\mathcal{S}_c (\mathbb{R}^m)$ can naturally be extended to a unital algebra with; $\mathcal{S}_{c,1} (\mathbb{R}^m):=\mathcal{S}_c (\mathbb{R}^m)\oplus \mathbb{C}$. It is obvious that for any $f\in \mathcal{S}_{c,1} (\mathbb{R}^m)$, $1 \star f=f \star 1=f$ if and only if;
\begin{eqnarray} \label {8}
\alpha(p,p)=\alpha(p,0)=0~,
\end{eqnarray}
\noindent for any $p\in \mathbb{R}^m$. Combining (\ref {7}) and (\ref {8}) leads to;
\begin{eqnarray} \label {9}
\alpha (0,p)=\alpha (0,-p)~,
\end{eqnarray}
\noindent for any $p\in \mathbb{R}^m$. Using (\ref {9}) it can also be shown that any translation-invariant product admits the trace property;
\begin{eqnarray} \label {10}
\int_{\mathbb{R}^m} f_1\star ...\star f_{k-1} \star f_k=\int_{\mathbb{R}^m} f_k\star f_1\star ...\star f_{k-1}~,
\end{eqnarray}
\noindent for any $k\in \mathbb{N}$ and for any set of $f_1,...,f_{k-1},f_k\in \mathcal{S}_c (\mathbb{R}^m)$.\\


 \par To characterize the translation-invariant star products effectively, 2-cocycles $\alpha$ should be categorized and classified appropriately. Conventionally, the 2-cocycles are studied in the setting of a cohomology theory [1, 2, 5, 7] usually referred to as $\alpha$-cohomology.  By definition [7] the $\alpha$-cohomology groups are the cohomology groups of the following complex;
\begin{eqnarray} \label {11}
\emph{\emph{C}}^0(\mathbb{R}^m)
\begin{array}{c}
  {\partial_0} \\
  {\longrightarrow} \\
  { }
\end{array}
\emph{\emph{C}}^1(\mathbb{R}^m)
\begin{array}{c}
  {\partial_1} \\
  {\longrightarrow} \\
  { }
\end{array}
...
\begin{array}{c}
  {\partial_{n-1}} \\
  {\longrightarrow} \\
  { }
\end{array}
\emph{\emph{C}}^n(\mathbb{R}^m)
\begin{array}{c}
  {\partial_{n}} \\
  {\longrightarrow} \\
  { }
\end{array}
...
\end{eqnarray}
with;
\begin{itemize}
  \item $\emph{\emph{C}}^0(\mathbb{R}^m):=\{0\}$,
  \item For $n=1$; $\emph{\emph{C}}^1(\mathbb{R}^m):=\{f\in C^\infty(\mathbb{R}^m)|f(0)=0\}$,
  \item For $n=2$; $\emph{\emph{C}}^2(\mathbb{R}^m):=\{f\in C^\infty(\mathbb{R}^m\times \mathbb{R}^m)|f(p,0)=f(p,p)=0; p\in \mathbb{R}^m\}$,
  \item For $n\geq3$; $\emph{\emph{C}}^n(\mathbb{R}^m)\subseteq C^\infty(\underbrace{\mathbb{R}^m\times ...\times\mathbb{R}^m}_{n- \emph{\emph{fold}}})$ consists of smooth functions $f$ with properties of $f(p_1,...,p_{n-1},0)=f(p_1,...,p_k,p,p,p_{k+1},...,p_{n-2})=0$, $k\leq {n-2}$, for any $p,p_1,...,p_{n-1}\in \mathbb{R}^m$,\\
\end{itemize}
\noindent and for the linear maps
\begin{eqnarray} \label {12}
\partial_n:\emph{\emph{C}}^n(\mathbb{R}^m)\longrightarrow \emph{\emph{C}}^{n+1}(\mathbb{R}^m)~,
\end{eqnarray}
\noindent usually denoted by $\partial$, defined by;
\begin{eqnarray} \label {13}
\partial_n f(p_0,...,p_n):=\varepsilon_n \sum _{i=0}^n f(p_0,...,p_{i-1},\hat{p _i},p_{i+1},...,p_n)+\varepsilon_n (-)^{n+1} f(p_0-p_n,...,p_{n-1}-p_n)~,
\end{eqnarray}
$f\in\emph{\emph{C}}^n(\mathbb{R}^m)$, with $\varepsilon_n=1$ for odd $n$ and $\varepsilon_n=i$ for $n$ even. One should note that; $\partial^2=\partial_n\circ\partial_{n-1}=0$ for any $n\in \mathbb{N}$.\\


 \par Conventionally the notation of $\alpha_1\thicksim\alpha_2$ is used for two $\alpha$-cohomologous $n$-cocycles $\alpha_1$ and $\alpha_2$. Also the cohomology class of $\alpha\in Ker\partial_n$ is shown by $[\alpha]$. Therefore, the $\alpha$-cohomolgy group, $H_\alpha^n (\mathbb{R}^m):=Ker\partial_n/Im\partial_{n-1}$, classifies $n$-cocycles differing in coboundary terms into the same equivalence classes. Now consider the translation-invariant products given by $\alpha\in C^\infty (\mathbb{R}^m\times\mathbb{R}^m)$ due to definition (\ref {6}). According to (\ref {7}), associativity of $\star$ is equivalent to $\partial\alpha=0$. Indeed, $H_\alpha^2 (\mathbb{R}^m)$ classifies all the translation-invariant quantization structures over $\mathcal{S}_{c,1} (\mathbb{R}^m)$ modulo the coboundary terms. It can be easily seen from (\ref {13}) that if $[\alpha]=0$ then $\alpha$ leads to a commutative star product $\star$, i.e.;
\begin{eqnarray} \label {14}
f\star g=g\star f~,
\end{eqnarray}
\noindent $f,g\in \mathcal{S}_{c,1} (\mathbb{R}^m)$. In [7] it has been proven that $\alpha$ leads to a commutative star product if and only if $[\alpha]=0$. Therefore, $\alpha_1\thicksim\alpha_2$ if and only if $\alpha_1-\alpha_2$ generates a commutative product.\\


 \par It can be shown that [7] there exists an algebraic version of Hodge theorem for $H_\alpha^2 (\mathbb{R}^m)$. More precisely, according to [7] for any given $\alpha$-cohomology class $[\alpha]\in H_\alpha^2 (\mathbb{R}^m)$, there exists a unique 2-cocycle, conventionally referred to as the harmonic form, which obeys the following properties;
\begin{eqnarray} \label {15}
\left\{
  \begin{array}{ll}
    \alpha(p,q)=-\alpha(p,p-q) \\
    \alpha(p,q)=\alpha(-p,-q)~~~~,  \\
    \alpha(p,q)=-\alpha(q,p)
  \end{array}
\right.
\end{eqnarray}
\noindent for any $p,q\in \mathbb{R}^m$.\\


 \par Specially for any 2-cocycle $\alpha$ its $\alpha$-cohomologous harmonic form, $\alpha_H$, is given by;
\begin{eqnarray} \label{16}
\alpha_H (p,q)=\frac{\alpha(p+q,q)-\alpha(p+q,p)}{2}~,
\end{eqnarray}
\noindent for any $p,q\in \mathbb{R}^m$.\\


 \par The set of all pure imaginary elements of $H_\alpha^2 (\mathbb{R}^m)$, denoted by
$H_{\alpha^*}^2 (\mathbb{R}^m)$, also defines a cohomology theory as a sub-theory of $\alpha$-cohomology, called the $\alpha^*$-cohomology, which also admits the Hodge theorem [7]. Particularly, $H_{\alpha^*}^2 (\mathbb{R}^m)$ classifies the complex translation-invariant products modulo the commutative ones. Consequently, according to the quantum equivalence theorem [7] due to (\ref{3}), $H_{\alpha^*}^2 (\mathbb{R}^m)$ classifies all the physically equivalent translation-invariant non-commutative versions of quantum field theories.\\


 \par For instance the Groenewold-Moyal star product, $\star_{G-M}$, and the Wick-Voros star product, $\star_{W-V}$, according to (\ref {6}) are respectively defined with 2-cocycles $\alpha_{G-M}(p,q)=iq^\mu \theta_{A \mu\nu} p^\nu$ and $\alpha_{W-V}(p,q)=\alpha_{G-M} (p,q)+q^\mu \theta_{S \mu\nu}(p-q)^\nu$, $p,q\in \mathbb{R}^m$, for $\theta_A$ an anti-symmetric and $\theta_S$ a symmetric real matrix. Therefore, it is seen that $\alpha_{G-M}$ and $\alpha_{W-V}$ are $\alpha^*$-cohomologous and belong to the same class of $H_{\alpha^*}^2 (\mathbb{R}^m)$ denoted by $[\alpha_{G-M}]$. Consequently, according to the quantum equivalence theorem, the Groenewold-Moyal and the Wick-Voros non-commutative versions of any quantum field theory lead to the same physics, the fact of which confirms the results of [1, 3, 9]. On the other hand, according to (\ref{15}) it is easily seen that $\alpha_{G-M}$ is a harmonic form. More precisely, any Groenewold-Moyal star product defines a particular class of $H_{\alpha^*}^2 (\mathbb{R}^m)$.\\

\section{Harmonic Forms and Groenewold-Moyal Star Products}
\setcounter{equation}{0}


 \par\indent In the last section it was finally shown that any given Groenewold-Moyal star product defines a particular class of $H_{\alpha^*}^2 (\mathbb{R}^m)$. Following this statement in this section it is shown that the converse is true, i.e. any arbitrary class of $H_{\alpha^*}^2 (\mathbb{R}^m)$ is uniquely characterized by a particular Groenewold-Moyal star product. To see this fact explicitly one initially needs to consider an arbitrary 2-cocycle $\alpha$ which satisfies the condition (\ref {7}). Showing the first and the second arguments of 2-cocycle $\alpha$ respectively by $z$ and $z'$, and then taking the partial derivative of (\ref {7}) with respect to $r^i$ at $r=0$, one finds that;
\begin{eqnarray} \label {17}
\frac{\partial\alpha}{\partial z'^i}(q,0)=\frac{\partial\alpha}{\partial z'^i}(p,0)-\frac{\partial\alpha}{\partial z^i}(p,q)-\frac{\partial\alpha}{\partial z'^i}(p,q)~,
\end{eqnarray}
\noindent $p,q\in \mathbb{R}^m$. Taking the partial derivative of (\ref {17}) with respect to $q^i$ leads to;
\begin{eqnarray} \label {18}
\frac{\partial^2\alpha}{\partial z'^j\partial z^i}(q,0)=-\frac{\partial^2\alpha}{\partial z'^i \partial z^j} (p,q)-\frac{\partial^2\alpha}{\partial z'^i \partial z'^j}(p,q)~,
\end{eqnarray}


\noindent $p,q\in \mathbb{R}^m$. Eventually, one needs to evaluate the partial derivative of (\ref {18}) with respect to $p^i$;
\begin{eqnarray} \label {19}
0=\frac{\partial^3\alpha}{\partial z^i \partial z'^j \partial z^k}(p,q)+\frac {\partial^3 \alpha}{\partial z^i \partial z'^j \partial z'^k}(p,q)~,
\end{eqnarray}
\noindent $p,q\in \mathbb{R}^m$. More precisely, it is seen that;
\begin{eqnarray} \label {20}
(\frac{\partial}{\partial z^k}+\frac{\partial}{\partial z'^k}) \frac{\partial^2 \alpha}{\partial z^i \partial z'^j}(p,q)=0~,
\end{eqnarray}
\noindent $p,q\in \mathbb{R}^m$. Using the coordinate transformation
\begin{eqnarray} \label {21}
\left\{
  \begin{array}{ll}
    z^{+i}=(z^i+z'^i)/\sqrt[]{2}~, \\
    z^{-i}=(z^i-z'^i)/\sqrt[]{2}~,
  \end{array}
\right.
\end{eqnarray}
\noindent one easily finds;
\begin{eqnarray} \label {22}
\frac{\partial}{\partial z^{+k}} \frac{\partial^2 \alpha}{\partial z^i \partial z'^j} (p,q)=0~,
\end{eqnarray}
\noindent $p,q\in \mathbb{R}^m$. Therefore;
\begin{eqnarray} \label {23}
\frac{\partial^2\alpha}{\partial z^i \partial z'^j}(p,q)={\sigma(\alpha)}_{ij}(p-q)~,
\end{eqnarray}
\noindent $p,q\in \mathbb{R}^m$.\\


 \par According to the last property of (\ref {15}), if $\alpha$ is harmonic it is anti-symmetric under the exchange of $p$ and $q$. Thus, for $\alpha$ a harmonic form one finds that;
\begin{eqnarray} \label {24}
{\sigma(\alpha)}_{ij}(p-q)=-{\sigma(\alpha)}_{ji}(q-p)~,
\end{eqnarray}
\noindent $p,q\in \mathbb{R}^m$. Moreover, if $\alpha$ is harmonic then by the second property of (\ref {15}) it is seen that;
\begin{eqnarray} \label {25}
{\sigma(\alpha)}_{ij}(p-q)={\sigma(\alpha)}_{ij}(q-p)~,
\end{eqnarray}
\noindent $p,q\in \mathbb{R}^m$. Hence, by (\ref {24}) and (\ref {25});
\begin{eqnarray} \label {26}
{\sigma(\alpha)}_{ij}(p-q)=-{\sigma(\alpha)}_{ji}(p-q)~,
\end{eqnarray}
\noindent $p,q\in \mathbb{R}^m$. Finally it can be seen that if $\alpha$ is harmonic then;
\begin{eqnarray} \label {27}
{\sigma(\alpha)}_{ij}(p-q)={\sigma(\alpha)}_{ij}(q)+\frac{\partial^2 \alpha}{\partial z'^i \partial z'^j}(p,p-q)~,
\end{eqnarray}
\noindent $p,q\in \mathbb{R}^m$, provided by the first property of (\ref {15}).\\
 \par According to (\ref {26}), ${\sigma(\alpha)}_{ij}$ is anti-symmetric under the exchange of indices $i$ and $j$. Therefore by (\ref {27}) one concludes that;
\begin{eqnarray} \label {28}
\frac{\partial^2\alpha}{\partial z^{'i}\partial z^{'j}}(p,q)=0~,
\end{eqnarray}
\noindent for any $p,q\in \mathbb{R}^m$, and consequently (\ref {27}) leads to;
\begin{eqnarray} \label {29}
{\sigma(\alpha)}_{ij}=\theta_{ij}\in \mathbb{C}~,
\end{eqnarray}
\noindent for any harmonic form $\alpha$. Eventually from (\ref {15}) and (\ref {29}) it is obvious that $\alpha$ is a harmonic form if and only if
\begin{eqnarray} \label {30}
\alpha(p,q)=p^i \theta_{ij} q^j~,
\end{eqnarray}
\noindent for any $p,q\in \mathbb{R}^m$, and for $\theta$ an anti-symmetric constant matrix. Therefore, by considering the pure imaginary harmonic forms, one concludes that $H_{\alpha^*}^2 (\mathbb{R}^m )$ is exactly the collection of Groenewold-Moyal star products. This lets one to characterize the cohomology groups $H_\alpha^2 (\mathbb{R}^m)$ and $H_{\alpha^*}^2 (\mathbb{R}^m )$ thoroughly with anti-symmetric $m \times m$ matrices. In fact, according to the Hodge theorem for $\alpha$-cohomology it is seen that;
\begin{eqnarray} \label {31}
H_\alpha^2 (\mathbb{R}^m)=\{\theta\in \mathbb{M}_{m \times m} (\mathbb{C})|\theta~ \emph{\emph{is anti-symmetric}}\}~.
\end{eqnarray}
 \par Therefore, $\emph{\emph{dim}}~H_\alpha^2 (\mathbb{R}^m )=m(m-1)$. Moreover, it is known [7] that $H_{\alpha^*}^2 (\mathbb{R}^m )$ is the collection of pure imaginary elements of $H_\alpha^2 (\mathbb{R}^m)$, thus;
\begin{eqnarray} \label {32}
H_{\alpha^*}^2 (\mathbb{R}^m )=\{\theta\in \mathbb{M}_{m\times m}(\mathbb{R})|\theta~\emph{\emph{is~anti-symmetric}}\}~.
\end{eqnarray}
 \par Consequently; $\emph{\emph{dim}}~H_{\alpha^*}^2 (\mathbb{R}^m)=m(m-1)/2$.\\
 \par In the last section it was shown that any given Groenewol-Moyal star product defines a particular class of $H_{\alpha^*}^2 (\mathbb{R}^m)$. Conversely, (\ref {30}) asserts that any class of $H_{\alpha^*}^2 (\mathbb{R}^m)$ is uniquely characterized by a particular Groenewold-Moyal star product. Therefore, one naturally concludes that any complex 2-cocycle $\alpha$ is $\alpha^*$-cohomologous to a particular Groenewold-Moyal 2-cocycle. Particularly (\ref {30}) shows that for any complex translation-invariant star product $\star_i$, there is a unique Groenewold-Moyal star product, say $\star_{i/G-M}$, such that;
\begin{eqnarray} \label {33}
\star_i\thicksim \star_{i/G-M}~.
\end{eqnarray}


 \par Using (\ref {33}) and the quantum equivalence theorem due to (\ref {3}) it is seen that for any general translation-invariant non-commutative quantum field theory there is a particular Gronwold-Moyal non-commutative quantum field theory with exactly the same effects and physical out-comings such as $n$-point functions and the scattering matrix. Then studying the Groenewold-Moyal non-commutative quantum field theories covers the whole domain of translation-invariant non-commutative quantum field theories.\\
 \par Translation-invariant star products also can be defined over the polynomials of coordinate functions. This leads to non-commutative structures of space-time. More precisely, the non-commutative structure of space-time due to 2-cocyle $\alpha$ is given by;
\begin{eqnarray} \label {34}
[x^i,x^j]_\star =x^i\star x^j-x^j\star x^i=\frac{\partial^2 \alpha}{\partial z^j \partial z^{'i}}(0,0)-\frac{\partial^2\alpha}{\partial z^i \partial z^{'j}}(0,0)~,
\end{eqnarray}
\noindent $i,j=1,...,m$, for $\star$ the translation-invariant star product induced by $\alpha$. It can be seen that if $\star$ is commutative or equivalently [7] if $\alpha=\partial\beta$ for 1-cochain $\beta$, then;
\begin{eqnarray} \label {35}
[x^i,x^j]_\star=0~,
\end{eqnarray}
\noindent $i,j=1,...,m$. Equality (\ref {35}) shows that the non-commutative structure of space-time is particularly given by the $\alpha^*$-cohomology class of the 2-cocycle.  Actually if $\star_1\thicksim\star_2$ then
\begin{eqnarray} \label {36}
[x^i,x^j]_{\star_1}=[x^i,x^j]_{\star_2}~,
\end{eqnarray}
\noindent $i,j=1,...,m$. Consequently the non-commutative structure of space-time due to 2-cocyle $\alpha$ can be precisely given by its $\alpha^*$-cohomologous harmonic form or more clearly by its $\alpha^*$-cohomologous Groenewold-Moyal star product. In fact, if $\star$ is induced by 2-cocycle $\alpha$ with $\alpha(p,q)\thicksim ip^i \theta_{ij} q^j$, $p,q\in \mathbb{R}^m$, $\theta_{ij}\in \mathbb{R}$, then;
\begin{eqnarray} \label {37}
[x^i,x^j]_=2i\theta_{ij}~,
\end{eqnarray}
\noindent $i,j=1,...,m$. Equality (\ref {37}) can be considered as the converse proposition for statement (\ref {36}). In fact, (\ref {37}) asserts that if (\ref {36}) holds for complex translation-invariant star products $\star_1$ and $\star_2$, then $\star_1\thicksim\star_2$. Consequently by (\ref {36}) and (\ref {37}) one naturally concludes that $\star_1\thicksim\star_2$ if and only if $\star_1$ and $\star_2$ lead to the same non-commutative structure of space-time. Therefore, $H_{\alpha^*}^2 (\mathbb{R}^m)$ classifies the non-commutative structures of space-time. One important consequence of this achievement with regard to the quantum equivalence theorem is that the non-commutative structure of space-time thoroughly explains the structure of quantum behaviors of non-commutative quantum field theories. More precisely, the only fundamental data through the quantum physics points of view is the non-commutative structure of space-time, but not the analysis of the star product. This fact was partly proved for Wick-Voros and Groenewold-Moyal non-commutative star products \cite{Lizzi, Galluccio, Vitale, Lizzi2}, but here it has been provided a general proof for all cases. In fact, (\ref {37}) can be considered as a modified version of Kontsevich's theorem \cite{Kontsevich} which asserts that there is a one to one correspondence between Poisson structures and equivalent star products over a smooth manifold, noting that any non-commutative structure of space-time is essentially a Poisson structure. But there is a particular difference between Kontsevich's theorem and equation (\ref {37}): There is considered no symmetry in the Kontsevich's theorem for equivalent star products, while here star products are classified with insistence on translation-invariance. Moreover, there is no attention to quantum behaviors via classification of star products in Kontsevich's theorem, while here the critical property of our classification is preserving the quantum effects due to quantum equivalence theorem \cite{Varshovinew}. \\


 \par On the other hand, by (\ref {30}) and (\ref {37}) one simply concludes that there is not any non-commutative translation-invariant star product on $\mathcal{S}_{c,1} (\mathbb{R}^m)$ which leads to commutative space-time. More precisely, there is no translation-invariant non-commutative star product on $\mathcal{S}_{c,1} (\mathbb{R}^m)$ which is commutative at the level of coordinate functions. Therefore, due to path integral formalism where the integration is taken over $\mathcal{S}_{c,1} (\mathbb{R}^m)$, commutative space-time never admits non-commutative translation-invariant quantum field theories.\\


\par One of the other important conclusions of (\ref {30}) and the quantum equivalence theorem due to (\ref {3}) is that the Grosse-Wulkenhaar approach [13, 14] and the method of $1/p^2$ [15] also work properly for any given translation-invariant non-commutative version of $\phi^4$ theory. On the other hand, it can similarly be concluded that any proposal for renormalizing the Groenewold-Moyal non-commutative gauge theories extends thoroughly to the collection of all translation-invariant non-commutative gauge theories. More precisely, since any given complex 2-cocycle $\alpha$ can be uniquely decomposed to $\alpha=\alpha_{G-M}+\partial\beta$ due to the Hodge theorem in $\alpha^*$-cohomology \cite{Varshovinew}, the regularization methods for Feynman diagrams in Groenewold-Moyal non-commutative quantum field theories work well for all translation-invariant quantum field theories due to canceling out the coboundary terms, such as $\partial\beta$, for internal momenta of loop calculations \cite{Varshovinew}.\\


 \par It has already been shown that the Groenewold-Moyal non-commutative versions of relativistic quantum field theories admit the Drinfeld's twist of Poincare invariance as a modified concept of relativity \cite{Galluccio, Lizzi2, Chaichian1, Chaichian2, Woronowicz}. More precisely, it has been shown that the algebra of $\mathcal{S}_{c,1} (\mathbb{R}^m)_{\star_{G-M}}$ is also an algebra in the category of $U(\mathfrak{P}_m)_{\chi_{G-M}}$-modules \cite{Kassel}, where $U(\mathfrak{P}_m)$ is the universal enveloping algebra of the Poincare Lie algebra $\mathfrak{P}_m$ for $\{M_{\mu,\nu}\}_{\mu,\nu=0}^{m-1}$ and $\{P_\mu\}_{\mu=0}^{m-1}$, respectively the Lorentz and translation Lie algebra generators, and $U(\mathfrak{P}_m)_{\chi_{G-M}}$ is its Drinfeld's twist due to counital Groenewold-Moyal 2-cocycle $\chi_{G-M}:=\exp(-i\hbar^{-2}~\theta^{\mu\nu}~P_\mu\bigotimes P_\nu)$ \cite{Majid}. Due to the Hodge decomposition theorem for $\alpha^*$-cohomology which uniquely splits any given 2-cocycle $\alpha$ to a Groenewold-Moyal 2-cocycle, $\alpha_{G-M}$, and a coboundary term, say $\partial\beta$, it can be easily seen that the algebra of $\mathcal{S}_{c,1} (\mathbb{R}^m)_{\star}$ is also an algebra in the category of $U(\mathfrak{P}_m)_{\chi}$-modules for $\chi:=\exp(\beta(\vec{P} \bigotimes 1)+\beta(1 \bigotimes \vec{P}))~\chi_{G-M}~\exp(-\beta(\vec{P} \bigotimes 1+1 \bigotimes \vec{P}))$, where $\star$ is generated by 2-cocycle $\alpha_{G-M}+\partial\beta$. Therefore, any translation-invariant non-commutative version of a quantum field theory admits the twisted Poincare symmetry (due to the Drinfeld's twist of Poincare universal enveloping algebra, $U(\mathfrak{P}_m)$, for counital 2-cocycle $\chi$) as a modified meaning of relativistic invariance. It would also be interesting to note that $\chi=\partial_+ \gamma~\chi_{G-M}~\partial_- \gamma$ for counital 1-cochain $\gamma=\exp(\beta(\vec{P}))$, and therefore $\chi$ and $\chi_{G-M}$ are cohomologous in the the second cohomology space of Poincare universal enveloping Hopf algebra $\mathcal{H}^2(U(\mathfrak{P}_m))$ \cite{Majid}, i.e.; the Drinfeld's twist of Poincare universal enveloping algebra $U(\mathfrak{P}_m)$ due to counital 2-cocycles $\chi_{G-M}$ and $\chi$ lead to isomorphic Hopf algebras. Thus, the cohomology space of $\mathcal{H}^2(U(\mathfrak{P}_m))$ also classifies translation-invariant quantum field theories with the same quantum behaviors. Moreover, it can be easily seen that $\mathcal{H}^2(U(\mathfrak{T}_m)) \cong  H^2_\alpha (\mathbb{R}^m)$ where $\mathcal{H}^2(U(\mathfrak{T}_m))$ is the second cohomology group of commutative universal enveloping Hopf algebra $U(\mathfrak{T}_m)$ for $m$-dimensional translation Lie algebra $\mathfrak{T}_m$ generated by $\{P_\mu\}_{\mu=0}^{m-1}$. In fact, the classification of translation-invariant non-commutative quantum field theories due to quantum equivalence theorem of $\alpha^*$-cohomology can also be worked out in the setting of cohomology spaces of quantum groups.\\


 \par Finally according to (\ref {30}) it can be seen that the star product due to 2-cocycle $\alpha$ is naturally reflected by a modified version of Weyl map [22]. More precisely if $\alpha(p,q)=ip^i \theta_{ij} q^j+\partial\beta(p,q)$, $p,q\in \mathbb{R}^m$, $\theta_ij\in \mathbb{R}$, $i,j=1,...,m$, for 1-cochain $\beta$, then the star product of $\star$ according to (\ref {6}) is particularly reflected by the Weyl-Wigner correspondence [22, 23] due to the following modified version of Weyl map;
\begin{eqnarray} \label {38}
\hat{f}=\int \frac{\emph{\emph{d}}^m p}{(2\pi)^m}  e^{i\sum_{i=1}^m p_i{\hat{x}}_i} \tilde{f}(p) e^{\beta(p)}~,
\end{eqnarray}
\noindent for $f\in C^\infty (\mathbb{R}^m)$, $\tilde{f}$ its Fourier transform and for $\hat{f}$ its corresponding operator with $[\hat{x}_i,\hat{x}_j]=i\theta_{ij}$, $1\leq i,j\leq m$. Particularly, (\ref {30}) asserts that any translation-invariant non-commutative star product is reflected by a modified version of Moyal-Weyl-Wigner quantization due to (\ref {38}).


\section{Conclusions}
\par In this article $\alpha^*$-cohomology was studied thoroughly and it was shown that in each cohomology class there is a unique 2-cocycle, the harmonic form [7], which generates a particular Groenewold-Moyal star product. According to [7] where it was shown that any two $\alpha^*$-cohomologous 2-cocycles lead precisely to two equivalent quantum field theories, i.e. two quantum field theories with exactly the same scattering matrix, a one to one correspondence between the collection of Groenewold-Moyal $\star$ products and the set of quantum equivalent translation-invariant non-commutative quantum field theories was then worked out. More precisely, in this article it was shown that for any general translation-invariant quantum field theory there is a unique Groenewold-Moyal non-commutative quantum field theory with the same scattering matrix. As a corollary one concludes that studding only the Groenewold-Moyal non-commutative quantum field theories covers thoroughly the whole domain of translation-invariant non-commutative quantum field theories. On the other hand, it was explicitly shown that the non-commutative structure of space-time entirely describes the quantum behaviors of translation-invariant non-commutative quantum field theories, the conjecture of which was never precisely proved before. Consequently, it was particularly proved that for a fixed quantum field theory two of its non-commutative versions with complex translation-invariant star products $\star_1$ and $\star?_2$ are quantum equivalent if and only if $\star_1$ and $\star_2$ lead to the same non-commutative structure for space-time. Moreover, it was then discussed that the Grosse-Wulkenhaar approach and the method of $1/p^2$ also work properly for any given translation-invariant non-commutative version of $\phi^4$ theory. As a conclusion, it was illustrated that any proposal for renormalizing the Groenewold-Moyal non-commutative gauge theories extends thoroughly to the collection of all translation-invariant non-commutative gauge theories. It was also shown that any translation-invariant (non-commutative) quantum field theory admits a modified structure of relativistic invariance via the twisted Poincare symmetry due to its star product. Finally it was precisely established that any given translation-invariant non-commutative star product is reflected by a modified version of Moyal-Weyl-Wigner quantization.

\section{Acknowledgments}
\par The author should say his special gratitude to M. M. Sheikh-Jabbari for use-full comments and fruit-full discussions. Moreover, my deepest thanks go to A. Shafiei Deh Abad for his kind considerations and his motivating ideas. Also I should confess that this article owes most of its appearance to S. Ziaee, whom my deepest regards goes to for many things.



\end{document}